\documentclass[aps,nofootinbib,longbibliography,prd,onecolumn]{revtex4}
\usepackage[a4paper,top=2cm,bottom=2cm,left=2cm,right=2cm]{geometry}
\usepackage[linktocpage=true,colorlinks,citecolor=blue,linkcolor=blue,urlcolor=blue]{hyperref}
\usepackage{graphicx}
\usepackage{xcolor}
\usepackage{amsmath,amssymb,amsfonts}
\usepackage{mathtools,mathrsfs}
\usepackage[applemac]{inputenc}
\usepackage{enumerate}
\usepackage{dsfont}
\usepackage{tabularx}
\usepackage{float}
\usepackage{comment}
\usepackage[default]{gillius}
\usepackage{MyCommands}
\usepackage[
 format = hang, 
 justification = RaggedRight, 
 textfont = it, 
 figurename = Figure,
 ]{caption}
\usepackage{setspace}
\usepackage[normalem]{ulem}
\usepackage[babel]{csquotes}

\def\Q{\mathbb{Q}}
\def\T{\mathbb{T}}
\def\M{\mathcal{M}}
\def\N{\mathcal{N}}
\def\C{\mathcal{C}}
\def\R{\mathcal{R}}
\def\Lie{\mathcal{L}}

\makeatletter
\renewcommand{\thetable}{\@arabic\c@table}
\makeatother

\begin{document}
\title{Revisiting Cosmologies in Teleparallelism}
\author{Fabio D'Ambrosio}\email{fabioda@phys.ethz.ch}
\author{Lavinia Heisenberg}\email{lavinia.heisenberg@phys.ethz.ch}
\author{Simon Kuhn}\email{simkuhn@phys.ethz.ch}
\affiliation{Institute for Theoretical Physics,
ETH Zurich, Wolfgang-Pauli-Strasse 27, 8093, Zurich, Switzerland}

\begin{abstract}
\noindent We discuss the most general field equations for cosmological spacetimes  for theories of gravity based on non-linear extensions of the non-metricity scalar and the torsion scalar. Our approach is based on a systematic symmetry-reduction of the metric-affine geometry which underlies these theories. While for the simplest conceivable case  the connection disappears from the field equations and one obtains the Friedmann equations of General Relativity, we show that in $f(\mathbb{Q})$ cosmology the connection generically modifies the metric field equations and that some of the connection components become dynamical. We show that $f(\mathbb{Q})$ cosmology contains the exact General Relativity solutions and also exact solutions which go beyond. In $f(\mathbb{T})$~cosmology, however, the connection is completely fixed and not dynamical.
\end{abstract}

\maketitle
\bigskip\bigskip
\hrule
\spacing{1.7}
\tableofcontents
\singlespacing
\bigskip
\hrule

\section{Introduction}\label{sec:Intro}\setcounter{equation}{0}
General relativity (GR) is the most successful theory of gravity so far. However,  difficulties such as singularities in cosmologies and black holes, observational tensions in the cosmological standard model, the need of dark energy, dark matter and the inflaton field, and the quantum nature of gravity, led to considering generalizations or reformulations of GR. In fact, Einstein's geometric formulation of gravity based on curvature is not unique. Following in his footsteps, if we fully embrace the geometrical character of gravity advocated by the equivalence principle, we realize that gravity can be geometrised in distinct but equivalent manners. These equivalent formulations form what is known as the trinity of GR~\cite{BeltranJimenez:2019, Heisenberg:2018}. One cornerstone of this trinity is constituted by the standard formulation of GR in terms of curvature. The other two formulations are known as Metric Teleparallelism (MT) and Symmetric Teleparallelism (ST), where one has not only the metric $g_{\mu\nu}$ --which describes the gravitational field-- as the dynamical field, but also an affine connection. The latter is postulated to be flat and metric-compatible  in the case of MT, while ST rests on the geometric postulate of a  flat and symmetric connection. In MT, one constructs the so-called torsion scalar $\T$ from the torsion tensor and subsequently introduces the Lagrangian density $\Lie=\sqrt{-g}\,\T$. Similarly, in ST one utilises the non-metricity tensor to define the non-metricity scalar $\Q$, leading to the ST Lagrangian density $\Lie=\sqrt{-g}\,\Q$. One can then show that the resulting actions are, up to a boundary term, equivalent to the Einstein-Hilbert action of GR. Since the latter depends only on the metric, the connection is not dynamical in neither MT nor ST, as it can be absorbed in a boundary term in the action. In particular, it neither enters the MT nor the ST metric field equations, and its own equations of motion are mere identities.\\
This changes drastically when one considers non-linear extensions of MT and ST, obtained by choosing $\Lie=\sqrt{-g}f(\T)$ and $\Lie=\sqrt{-g}f(\Q)$, respectively, where $f$ is an arbitrary function. Now one can no longer absorb the connection into a boundary term, and it can thus potentially become dynamical or influence the metric equations of motion, and thus the gravitational field. This can lead to the propagation of new degrees of freedom and  consequently to solutions which go beyond standard GR.\\
In this paper we analyze cosmological field equations in $f(\T)$ and $f(\Q)$ gravity. In particular, we explore whether the most general metric-affine geometries which are compatible with the cosmological principle of homogeneity and isotropy, as well as with the geometric postulates of MT and ST, lead to propagating degrees of freedom stemming from the connection. To that end, we perform a systematic symmetry-reduction of the metric and the connection. In a second step, we subject the connection to the geometric postulates of MT and ST separately. This allows us to derive the most general field equations for cosmology in non-linear extensions of teleparallel theories. Our analysis is in this sense fully general and complete.

Interesting cosmological solutions in $f(\Q)$ gravity have already been investigated in a multitude of studies~\cite{Jimenez:2019, Barros:2020, Ayuso:2020, Frusciante:2021, Anagnostopoulos:2021, Arora:2021, Bahamonde:2021, Esposito:2021, Dimakis:2021, Atayde:2021, Hohmann:2021}. However, in the majority of these works the connection was forced to trivialize. In this work, we explicitly show how the connection can become truly dynamical in non-linear extensions of the non-metricity scalar, leading to cosmological solutions which go beyond GR. In the context of spherically symmetric backgrounds in teleparallel gravity, it was recently shown that lifting the restriction of a trivial connection has far-reaching consequences for the existence of beyond-GR solutions~\cite{DAmbrosio:2021}.

The paper is structured as follows. In section~\ref{sec:Introduction} we briefly introduce $f(\T)$ and $f(\Q)$ gravity and present their respective field equations. Most of this section is based on existing literature~\cite{BeltranJimenez:2017,Jimenez:2018} and the main purpose is to fix notations and conventions. In section~\ref{sec:SymmReduction} we implement the cosmological symmetries for the metric and the connection, following the procedure discussed in~\cite{Hohmann:2019, Hohmann:2019nat}. The symmetry reduced metric and connection are then used in  subsection~\ref{ssec:Connections} to implement the geometric postulates of MT and ST, respectively. Section~\ref{sec:FieldEquations} is then dedicated to deriving the most general  cosmological field equations and to comparing the $f(\T)$ and $f(\Q)$ cosmologies in a broad sense. Specific examples are subsequently discussed in section~\ref{sec:Examples}. In this section we explicitly show that $f(\Q)$ contains the exact cosmological solutions of GR and that it also gives rise to beyond-GR solutions which could be of physical interest. We bring our work to a conclusion in section~\ref{sec:Conclusion} with a brief summary and an outlook on future work. Throughout this paper we use natural units, $c=8\pi G=1$.\\

At the time of writing, it came to our attention that a similar work~\cite{Hohmann:2021}, also dedicated to $f(\Q)$ cosmology and other ST extensions, was posted on arXiv. In particular, we obtain the same results for the connections and equations of motion for $f(\Q)$ cosmology. Furthermore, our work can be regarded as a natural successor to the recent analysis concerning spherically symmetric and stationary solutions in $f(\Q)$ and $f(\T)$ gravity which has been put forward in~\cite{DAmbrosio:2021}.

\section{$f(\Q)$ and $f(\T)$ Gravity}\label{sec:Introduction}\setcounter{equation}{0}
Both, MT and ST are formulated in terms of a metric-affine geometry $(\M,g_{\mu\nu},\Gamma^\alpha{}_{\mu\nu})$, where $\M$ is a smooth four-dimensional  manifold, $g_{\mu\nu}$ a symmetric, non-degenerate metric tensor of signature $(-,+,+,+)$, and $\Gamma^\alpha{}_{\mu\nu}$ denotes the components of an affine connection. The latter defines the covariant derivative of vector fields and 1-forms, which are respectively given by
\begin{align}
    \nabla_\alpha V^\mu&=\partial_\alpha V^\mu+\Gamma^\mu{}_{\alpha\nu}V^\nu\notag\\
    \nabla_\alpha \omega_\mu&=\partial_\alpha \omega_\mu-\Gamma^\nu{}_{\alpha\mu}\omega_\nu.
\end{align}
There are three geometric objects which can be constructed from the connection and the metric, and which capture different aspects of the metric-affine geometry; The Riemann, or curvature, tensor $R^\alpha{}_{\beta\mu\nu}$, the torsion tensor $T^\alpha{}_{\mu\nu}$, and the non-metricity tensor $Q_{\alpha\mu\nu}$, defined by\footnote{We define symmetrization and skew-symmetrization of a tensor $T_{\mu\nu}$ by $2T_{(\mu\nu)}:=T_{\mu\nu}+T_{\nu\mu}$ and $2T_{[\mu\nu]}:=T_{\mu\nu}-T_{\nu\mu}$, respectively.}
\begin{align}
    R^\alpha{}_{\beta\mu\nu}&:= 2\partial_{[\mu}\Gamma^\alpha{}_{\nu]\beta}+2\Gamma^\alpha{}_{[\mu|\lambda|}\Gamma^\lambda{}_{\nu]\beta}\notag\\
    T^\alpha{}_{\mu\nu}&:= 2\Gamma^\alpha{}_{[\mu\nu]}\notag\\
    Q_{\alpha\mu\nu}&:= \nabla_\alpha g_{\mu\nu}=\partial_\alpha g_{\mu\nu}-2\Gamma^\lambda{}_{\alpha(\mu}g_{\nu)\lambda}.
\end{align}
Observe that the Riemann and the torsion tensors only depend on the connection, while the non-metricity tensor also involves the metric in its definition. As is known, these three tensors can be used to define different Lagrangian densities which lead to distinct but equivalent formulations of GR -- This is the so-called trinity of GR~\cite{BeltranJimenez:2019}. 

We recall that the standard, or curvature, formulation of GR is recovered by postulating that the metric-affine geometry is (a) torsionless ($T^\alpha{}_{\mu\nu} = 0$) and (b) metric-compatible ($Q_{\alpha\mu\nu} = 0$). It is well-known that these postulates immediately imply that the connection is uniquely given by the Levi-Civita connection $\left\{\alpha\atop{\mu\nu}\right\}$. This connection is completely determined by the metric and is explicitly given by
\begin{equation}
    \Gamma^\alpha{}_{\mu\nu}=\left\{{\alpha\atop{\mu\nu}}\right\}=\frac{1}{2}g^{\alpha\beta}\left(2\partial_{(\mu}g_{\nu)\beta}-\partial_\beta g_{\mu\nu}\right).
\end{equation}
The action functional of GR can then be written as
\begin{equation}
   	\mathcal S_\text{GR}[g] =\int_\M \dd^4x\,\left( \sqrt{-g}\, \R+\lambda_\alpha{}^{\mu\nu}T^\alpha{}_{\mu\nu}+\mu^{\alpha\mu\nu}Q_{\alpha\mu\nu}\right)+\mathcal S_\textsf{M},
\end{equation}
where $\lambda$ and $\mu$ are tensor densities which act as Lagrange multipliers to enforce the vanishing of torsion and non-metricity, and $g:= \det(g_{\mu\nu})$. The scalar $\R$ is the Ricci scalar of the Levi-Civita connection:
\begin{equation}
\R := \left.R^{\alpha}{}_{\mu\alpha\nu}g^{\mu\nu}\right|_{\Gamma^\alpha{}_{\mu\nu}=\left\{{\alpha\atop{\mu\nu}}\right\}}.
\end{equation}
There are two additional formulations one can obtain by employing different postulates on the metric-affine geometry: The Teleparallel Equivalent of GR (TEGR) is defined by demanding that the geometry is metric-compatible and flat (i.e., $Q_{\alpha\mu\nu} = 0$ and $R^\alpha{}_{\beta\mu\nu} = 0$), while the Symmetric Teleparallel Equivalent of GR (STEGR) follows when one imposes the conditions that the geometry is torsionless and flat (i.e, $T^\alpha{}_{\mu\nu} = 0$ and $R^\alpha{}_{\beta\mu\nu} = 0$).

Let us first discuss TEGR and then turn to STEGR. It is easy to show that in TEGR the connection is not completely fixed, but rather it can be chosen to be of the form
\begin{equation}
    \Gamma^\alpha{}_{\mu\nu}=(\Lambda^{-1})^\alpha{}_{\beta}\partial_\mu \Lambda^\beta{}_{\nu},
\end{equation}
where the matrix $\Lambda\in GL(4,\mathbb{R})$ is arbitrary. This form of the connection ensures that it satisfies the requirement of metric-compatibility and flatness. Moreover the action functional can be written as
\begin{equation}
    \mathcal S_\text{TEGR}[g, \Gamma] =\int_\M \dd^4x\,\left(\sqrt{-g}\, \T+\lambda_\alpha{}^{\beta\mu\nu}R^\alpha{}_{\beta\mu\nu}+\mu^{\alpha\mu\nu}Q_{\alpha\mu\nu}\right)+ \mathcal S_\textsf{M},
\end{equation}
where $\lambda$ and $\mu$ are again tensor densities which act as Lagrange multipliers to ensure the vanishing of curvature and non-metricity. The torsion scalar $\T$ is defined as
\begin{equation}
    \T := -\frac{1}{4}T_{\alpha\mu\nu}T^{\alpha\mu\nu}-\frac{1}{2}T_{\alpha\mu\nu}T^{\mu\alpha\nu}+T_\alpha T^\alpha,
\end{equation}
where
\begin{equation}
	T_\alpha := T^\mu{}_{\alpha\mu}.
\end{equation}
At first glance it may seem like the theory possesses a dynamical connection. However, one can prove the identity
\begin{equation}\label{eq:TEGRIdentity}
    \T - \R - 2\mathcal{D}_\mu T^\mu =0,
\end{equation}
where $\R$ is again the Ricci scalar with respect to the Levi-Civita connection and $\mathcal D_\mu$ is the covariant derivative operator with respect to the same Levi-Civita connection. This identity shows that $\mathcal S_\text{TEGR}[g, \Gamma]=\mathcal S_\text{GR}[g]$ up to a boundary term. Hence, STEGR and GR possess the same field equations for the metric while the field equations for the connection reduce to trivial identities.\\

Let us now turn to STEGR. In this case, one can show that a connection which satisfies the requirements of torsion-freeness and flatness can be written as
\begin{equation}
    \Gamma^\alpha{}_{\mu\nu} = \frac{\partial x^\alpha}{\partial\zeta^\lambda}\partial_\mu\partial_\nu\zeta^\lambda,
\end{equation}
where $\xi:\mathcal M\to\mathcal M$ is an arbitrary diffeomorphism (and $\xi^\mu$ simply denotes its components in $\M$). In particular, one realizes that it is always possible to choose a gauge in which the connection vanishes globally, $\Gamma^\alpha{}_{\mu\nu} = 0$. This is the so-called coincident gauge~\cite{BeltranJimenez:2017}. For the action functional one writes
\begin{equation}
    \mathcal S_\text{ST}[g, \Gamma] = \int_\M \dd^4x\,\left(\sqrt{-g}\, \Q+\lambda_\alpha{}^{\beta\mu\nu}R^\alpha{}_{\beta\mu\nu}+\mu_\alpha{}^{\mu\nu}T^\alpha{}_{\mu\nu}\right)+\mathcal S_\textsf{M},
\end{equation}
where the tensor densities $\lambda$ and $\mu$ enforce the vanishing of curvature and torsion, and the non-metricity scalar $\Q$ is defined as
\begin{equation}
    \Q := -\frac14\, Q_{\alpha\beta\gamma}Q^{\alpha\beta\gamma} + \frac12\, Q_{\alpha\beta\gamma}Q^{\beta\alpha\gamma}	 + \frac14\, Q_\alpha Q^{\alpha} - \frac12 \, Q_\alpha \bar{Q}^\alpha,
\end{equation}
with the two independent non-metricity traces
\begin{align}
	Q_\alpha &:= Q_{\alpha\mu}{}^\mu & \text{and} & & \bar{Q}_\alpha &:= Q_{\mu\alpha}{}^\mu.
\end{align}
Just as for TEGR, the STEGR action seems to lead to a dynamical connection. However, a similar identity holds which allows us to re-express the non-metricity scalar in terms of the Ricci scalar and a total derivative term:
\begin{equation}\label{eq:STIdentity}
    \Q - \R - \mathcal{D}_\mu(Q^\mu-\bar Q^\mu) =0.
\end{equation}
This demonstrates that $\mathcal S_\text{STEGR}[g, \Gamma]=\mathcal S_\text{GR}[g]$ up to a boundary term. Hence, STEGR and GR share the same metric field equations while the field equations for the connection reduce once more to mere identities.\\

This brief review of TEGR and STEGR fixes our conventions and notations, and it sets the stage for the rest of the paper. In fact, what we are interested in are non-linear extensions of these two theories. Namely, $f(\Q)$ and $f(\T)$ gravity. These extensions are defined by the action functionals
\begin{align}
    \mathcal S_{f(\T)}[g, \Gamma] &:=\int_\M \dd^4x\,\left(\sqrt{-g}\, f(\T)+\lambda_\alpha{}^{\beta\mu\nu}R^\alpha{}_{\beta\mu\nu}+\mu^{\alpha\mu\nu}Q_{\alpha\mu\nu}\right)+\mathcal S_\text{M}\notag\\
    \mathcal S_{f(\Q)}[g, \Gamma] &:=\int_\M \dd^4x\,\left(\sqrt{-g}\, f(\Q)+\lambda_\alpha{}^{\beta\mu\nu}R^\alpha{}_{\beta\mu\nu}+\mu_\alpha{}^{\mu\nu}T^\alpha{}_{\mu\nu}\right)+\mathcal S_\text{M},
\end{align}
where $f$ is an arbitrary function solely subjected to the condition that $f'\neq 0$\footnote{Here and in the sequel, a prime on $f$ will always denote a derivative with respect to its argument and not a derivative with respect to a coordinate.}. In both cases, there is no identity analogous to~\eqref{eq:TEGRIdentity} or~\eqref{eq:STIdentity} and the connection can therefore not simply be absorbed into a boundary term. Rather, the connection field equations will become non-trivial and the connection can therefore propagate its own degrees of freedom. These theories are therefore also genuine extensions of GR and, moreover, potentially different from $f(\R)$ gravity.

In the sequel, we study cosmological spacetimes in the context of $f(\T)$ and $f(\Q)$ gravity. To that end, we choose a vanishing hypermomentum~\cite{Jimenez:2018} and we assume the energy-momentum tensor to be given by
\begin{equation}
    T^\mu{}_{\nu} = \text{diag}\left(\rho,-p,-p,-p\right),
\end{equation}
where $\rho$ denotes the energy density and $p$ the pressure. Under the assumption of a vanishing hypermomentum, one can show that the metric and connection field equations of $f(\T)$ gravity can be written as~\cite{Jimenez:2018}
\begin{align}\label{eq:MTFieldEqs}
    \M_{\mu\nu} := (\nabla_\alpha+T_\alpha)[S_{(\mu\nu)}{}^{\alpha}f'(\T)]+f'(\T)t_{\mu\nu}-\frac{1}{2}f(\T)g_{\mu\nu} &= T_{\mu\nu} \notag\\
    \mathcal{C}_{\alpha\beta}:=-(\nabla_\mu+T_\mu)\left[\frac{\sqrt{-g}}{2}f'(\T) S_{[\alpha}{}^{\mu}{}_{\beta]} \right]&=0,
\end{align}
where $\M_{\mu\nu}$ are the metric field equations and $\C_{\alpha\beta}$ represent the connection field equations. Moreover, we have made use of the definitions
\begin{align}
    S_\alpha{}^{\mu\nu} &:=\frac{\partial \T}{\partial T^\alpha{}_{\mu\nu}}=-\frac{1}{2}T_\alpha{}^{\mu\nu}-{T^{[\mu}}_{\alpha}{}^{\nu]}-2\delta_\alpha{}^{[\mu}T^{\nu]}\notag\\
    t_{\mu\nu} &:=\frac{1}{2}S_\mu{}^{\alpha\beta}T_{\nu\alpha\beta}-T^{\alpha\beta}{}_{\mu}S_{\alpha\beta\nu}.
\end{align}
We will also need the field equations of $f(\Q)$ gravity, which can be written as~\cite{Jimenez:2018}
\begin{align}\label{eq:FieldEquations}
	\M_{\mu\nu} := \frac{2}{\sqrt{-g}}\nabla_\alpha\left[\sqrt{-g}P^\alpha{}_{\mu\nu} f'(\Q)\right] + f'(\Q) q_{\mu\nu} -\frac12 f(\Q) g_{\mu\nu} &= T_{\mu\nu} \notag\\
	\mathcal C_\alpha := \nabla_\mu\nabla_\nu\left(\sqrt{-g}\,f'(\Q) P^{\mu\nu}{}_{\alpha}\right) &= 0,
\end{align}
where $\M_{\mu\nu}$ and $\C_\alpha$ represent again the metric and connection field equations, respectively. We have also introduced the following tensors:
\begin{align}
	P^\alpha{}_{\mu\nu} &:= \frac12 \PD{\Q}{Q_\alpha{}^{\mu\nu}} = -\frac14 Q^\alpha{}_{\mu\nu} + \frac12 Q_{(\mu}{}^{\alpha}{}_{\nu)} +\frac14 g_{\mu\nu}Q^\alpha -\frac14 \left(g_{\mu\nu} \bar{Q}^\alpha + \delta^\alpha{}_{(\mu} Q_{\nu)}\right)\notag\\
	q_{\mu\nu} &:= \PD{\Q}{g^{\mu\nu}} = P_{(\mu|\alpha\beta}Q_{\nu)}{}^{\mu\nu} - 2P^{\alpha\beta}{}_{(\nu} Q_{\alpha\beta|\mu)}.
\end{align}
Before concluding this review section, we note for later reference that the metric field equations of both, $f(\T)$ and $f(\Q)$ gravity can be written in a very neat form. For $f(\T)$ gravity one finds
\begin{equation}\label{eq:MTSimpleMetEq}
    f'(\T)G_{\mu\nu}-\frac{1}{2}g_{\mu\nu}(f(\T)-f'(\T)\T)+f''(\T)S_{(\mu\nu)}{}^\alpha\partial_\alpha \T = T_{\mu\nu},
\end{equation}
while for $f(\Q)$ gravity one obtains the equations
\begin{equation}\label{eq:RewrittenMetricFieldEq}
	f'(\Q) G_{\mu\nu}  - \frac12 g_{\mu\nu} (f(\Q)-f'(\Q)\Q) + 2 f''(\Q) P^\alpha{}_{\mu\nu} \partial_\alpha\Q  = T_{\mu\nu}.
\end{equation}
In both cases, $G_{\mu\nu}$ stands for the Einstein tensor with respect to the Levi-Civita connection.

\section{Symmetry-Reduction of the Metric and the Connection}\label{sec:SymmReduction}\setcounter{equation}{0}
Our goal is to study cosmological spacetimes in the context of $f(\Q)$ and $f(\T)$ gravity without making any explicit choices or ``guesses'' for the connections. Rather, we want to use the most general forms of the metric and the connection which are compatible with the cosmological principles of homogeneity and isotropy, and which are compatible with the geometric postulates of $f(\Q)$ and $f(\T)$ gravity, respectively. To that end, we follow a similar strategy as the one employed in~\cite{DAmbrosio:2021} in the context of black holes in $f(\Q)$ and $f(\T)$ gravity. That is, we perform a systematic symmetry-reduction of the metric, the connection, and the field equations. We note that this has recently also been achieved in~\cite{Hohmann:2021}.

In order to perform the symmetry-reduction, we first choose a chart\footnote{Cosmological singularities could in principle restrict time $t$ to a (half)interval $\mathcal{I}\subset\mathbb{R}$.} $(t, r, \theta, \phi)\in \mathbb R\times\mathbb R_{>0}\times[0,\pi]\times[0,2\pi)$. To implement the requirement of homogeneity and isotropy, we demand that the metric and the connection are invariant under rotations and spatial translations. Mathematically, this amounts to demanding that the Lie derivatives of the metric and the connection vanish~\cite{Hohmann:2019nat} with respect to the following Killing vector fields
\begin{align}\label{eq:KVFs}
	\zeta_{\mathcal R_x} &=  \sin\phi\,\partial_\theta +\frac{\cos\phi}{\tan\theta}\,\partial_\phi, &  \zeta_{\mathcal{T}_x} &= \chi\,\sin\theta\cos\phi\, \partial_r+\frac{\chi}{r}\cos\theta\cos\phi\,\partial_\theta-\frac{\chi}{r}\frac{\sin\phi}{\sin\theta}\partial_\phi\notag\\
	\zeta_{\mathcal R_y} &=  -\cos\phi\,\partial_\theta + \frac{\sin\phi}{\tan\theta}\,\partial_\phi, & \zeta_{\mathcal{T}_y} &= \chi\,\sin\theta\sin\phi\, \partial_r+\frac{\chi}{r}\cos\theta\sin\phi\,\partial_\theta+\frac{\chi}{r}\frac{\cos\phi}{\sin\theta}\partial_\phi\notag\\
	\zeta_{\mathcal R_z} &=  -\partial_\phi, & \zeta_{\mathcal{T}_z} &= \chi\,\cos\theta\, \partial_r-\frac{\chi}{r}\sin\theta\,\partial_\theta,
\end{align}
where $\zeta_{\R_i}$ generate rotations and $\zeta_{\mathcal T_i}$ generate translations and where we have introduced the notation $\chi:=\sqrt{1-k\, r^2}$ in order to compactify some of the expressions which will follow. It turns out that it suffices~\cite{Hohmann:2019} to consider only $\zeta_{\R_i}$ and $\zeta_{\T_1}$, as the remaining two Killing vectors can be obtained by taking Lie brackets of these four. The symmetry conditions we have to study can then be written as
\begin{align}
    \Lie_\zeta g_{\mu\nu} &\overset{!}{=} 0 & \text{and} & &\Lie_\zeta \Gamma^\alpha{}_{\mu\nu} &\overset{!}{=}0.
\end{align}
Observe that these conditions provide use with equations which are linear in the connection and in the metric and we will thus find unique solutions. In the next two subsection, we will explicitly implement the symmetry conditions.

\subsection{Symmetry-Reduction of the Metric}
We begin with a generic metric $g_{\mu\nu}$ which has ten components which are all functions of $(t,r,\theta,\phi)$. Homogeneity and isotropy are then implemented by demanding that
\begin{equation}
    \Lie_\zeta g_{\mu\nu}=\zeta^\lambda\partial_\lambda g_{\mu\nu}+\partial_\mu\zeta^\lambda g_{\lambda\nu}+\partial_\nu\zeta^\lambda g_{\mu\lambda} \overset{!}{=}0,
\end{equation}
with respect to the rotational and translation Killing vector fields introduced in~\eqref{eq:KVFs}. The resulting equations are simple but quite numerous. Since the result of this particular symmetry-reduction is well-known, we simply state it without derivation:
\begin{equation}\label{eq:SymmetricMetric}
    \dd s^2=g_{\mu\nu}\dd x^\mu \dd x^\nu = g_{tt}(t)\,\dd t^2+g_{rr}(t)\left[\frac{\dd r^2}{\rchi}+r^2\left(\dd\theta^2+\sin^2\theta\,\dd\phi^2\right)\right].
\end{equation}
Hence, we have reduced the initial ten components of the metric to only two independent functions of $t$, namely $g_{tt}$ and $g_{rr}$. Moreover, the metric is diagonal and the parameter $k\in\mathbb R$ defines the spatial curvature. For $k=0$ one obtains flat spatial sections, while $k>0$ and $k<0$ describe spherical and hyperbolic sections, respectively.\\
\\
It is well-known that the component $g_{tt}$ is unphysical. A simple coordinate transformation
\begin{equation}
    \tilde t:=\int^t\,\sqrt{-g_{tt}(\tau)}\,\dd \tau
\end{equation}
transforms the line element to
\begin{equation}
    \dd s^2=-\dd\tilde t^2+g_{rr}(\tilde t)\left(\frac{\dd r^2}{\rchi}+r^2(\dd\theta^2+\sin^2\theta\,\dd\phi^2)\right).
\end{equation}
One can thus always set $g_{tt}=-1$. However, for the sake of generality, we will   keep $g_{tt}$ arbitrary. Similarly, for $k\neq 0$, one can rescale $r$  by
\begin{equation}
    \tilde r:=\sqrt{|k|}\,r
\end{equation}
accompanied by a rescaling $g_{rr}\to g_{rr}/|k|$. This would lead to a line element of the form~\eqref{eq:SymmetricMetric} but with $k$ now only takes the values $\{-1,0,1\}$. However, also in this case, we will keep $k$ general and always assume  $k\in \mathbb{R}$.

\subsection{Symmetry-Reduction of the Connection}
Let us now turn to the connection. We do not impose any geometric postulates on the connection at this stage. That is, we start with a generic connection which has $64$ components, all of which are function of $(t, r, \theta, \phi)$, and we do not demand flatness, metric-compatibility, nor torsion-freeness. We only demand that the connection respects homogeneity and isotropy, which is tantamount to demanding that
\begin{equation}
    \Lie_\zeta \Gamma^\alpha{}_{\mu\nu} = \zeta^\lambda\Gamma^\alpha{}_{\mu\nu}-\partial_\lambda\zeta^\alpha\Gamma^\lambda{}_{\mu\nu}+\partial_\mu\zeta^\lambda\Gamma^\alpha{}_{\lambda\nu}+\partial_\nu\zeta^\lambda\Gamma^\alpha{}_{\mu\lambda}+\partial_\mu\partial_\nu\zeta^\alpha \overset{!}{=} 0
\end{equation}
with respect to the Killing vector fields~\eqref{eq:KVFs}. Just as for the metric, this leads to simple but even more numerous equations. We therefore simply state the final result, which can be concisely formulated in terms of matrices:
\begin{align}\label{eq:SymmetricConnection}
\Gamma^t{}_{\mu\nu}&=\left(
\begin{array}{cccc}
 C_1 & 0 & 0 & 0 \\
 0 & \frac{C_2}{\chi^2} & 0 & 0 \\
 0 & 0 &  C_2\,r^2 & 0 \\
 0 & 0 & 0 &  C_2\,r^2 \sin^2\theta \\
\end{array}
\right),& \Gamma^r{}_{\mu\nu}&=\left(
\begin{array}{cccc}
 0 & C_3 & 0 & 0 \\
 C_4 & \frac{k\, r}{\chi} & 0 & 0 \\
 0 & 0 & -r\,\chi  & -C_5\,r^2\, \chi \, \sin\theta \\
 0 & 0 & C_5\,r^2\,\chi\, \sin\theta & -r\,\chi  \sin^2\theta \\
\end{array}
\right)\notag\\
\Gamma^{\theta}{}_{\mu\nu}&=\left(
\begin{array}{cccc}
 0 & 0 & C_3 & 0 \\
 0 & 0 & \frac{1}{r} & \frac{C_5\, \sin\theta}{\chi} \\
 C_4 & \frac{1}{r} & 0 & 0 \\
 0 & -\frac{C_5\, \sin\theta}{\chi} & 0 & -\cos\theta\, \sin\theta \\
\end{array}
\right), &
\Gamma^{\phi}{}_{\mu\nu}&=\left(
\begin{array}{cccc}
 0 & 0 & 0 & C_3 \\
 0 & 0 & -\frac{C_5\, \csc\theta}{\chi} & \frac{1}{r} \\
 0 & \frac{C_5\, \csc\theta}{\chi} & 0 & \cot\theta \\
 C_4 & \frac{1}{r} & \cot\theta & 0 \\
\end{array}
\right).
\end{align}
This symmetry reduced connection agrees with the one found in~\cite{Hohmann:2019}. The originally $64$ components of the connection have been reduced to only five arbitrary functions of $t$, namely the functions $C_i$ with $i\in\{1,2,3,4,5\}$, and a few components which are completely expressed by known elementary functions. It is possible to discuss a special case already at this point: Observe that when the spatial sections are flat, i.e., $k=0$, and when one chooses $C_i=0$ for all~$i$, then one obtains the connection in coincident gauge expressed in spherical coordinates\footnote{This means that when applying the coordinate transformation that transforms Cartesian to spatially spherically coordinates to the coincident gauge $\Gamma^\alpha{}_{\mu\nu}=0$ one obtains $\Gamma^s$.}. Hence, this connection is then trivially torsionless and flat. We call this connection the spherical connection and denote it by $\Gamma^s$. This connection is widely used in the ST and $f(\Q)$ cosmology literature~\cite{BeltranJimenez:2017,Jimenez:2018,Mandal:2020}.

\section{Symmetric Teleparallel and Metric Teleparallel Connections}\label{ssec:Connections}\setcounter{equation}{0}
In the previous section we have derived the most general metric and connection which respect the cosmological principles of homogeneity and isotropy. However, we have neither implemented the geometric postulates of $f(\Q)$ gravity nor the postulates of $f(\T)$ gravity. The implementation of these postulates will be the subject of this section and we start with implementing the flatness condition, $R^\alpha{}_{\beta\mu\nu}= 0$, since this is common to both theories. Afterward, we specialize the homogeneous, isotropic, and flat connection to $f(\T)$ (cf. subsection~\ref{ssec:MetricCompatibility}) and $f(\Q)$ (cf. subsection~\ref{ssec:ZeroTorsion}) separately.\\
\\
To start with, we plug the connection~\eqref{eq:SymmetricConnection} into the condition $R^\alpha{}_{\beta\mu\nu}\overset{!}{=} 0$. Since the Riemann tensor is quadratic in the connection and linear in its first derivatives, we find non-linear algebraic equations as well as non-linear first order differential equations. Indeed, one finds that the flatness condition reduces to the following equations:
\begin{align}\label{eq:Riemann1}
	\dot C_5&=0 & k+C_2C_4-C_5^2&=0\notag\\
    C_1C_4-C_3C_4-\dot C_4&=0 & C_2C_5&=0\notag\\
    C_1C_2-C_2C_3+\dot C_2 &=0 & C_4C_5&=0,
\end{align}
where a dot stands for a derivative with respect to $t$. Observe that the first equation in~\eqref{eq:Riemann1} is solved by $C_5=c$, where $c\in\mathbb R$ is a constant. The remaining equations are non-linear, leading to many solution sets for $C_1$, $C_2$, $C_3$, and $C_4$. We postpone the discussion of these solutions to after we have implemented the metric-compatibility condition for $f(\T)$ gravity and the condition of vanishing torsion for $f(\Q)$ gravity, as these conditions will have an effect on the connection and consequently also impact the above flatness conditions.

\subsection{Implementing Metric-Compatibility for the $f(\T)$ Connection}\label{ssec:MetricCompatibility}
In order to implement $Q_{\alpha\mu\nu}\overset{!}{=} 0$ we use the metric~\eqref{eq:SymmetricMetric}, the connection~\eqref{eq:SymmetricConnection}, and we use $C_5=c$. We then find the following equations
\begin{align}\label{eq:ZeroNonMet}
    C_1-\frac{\dot g_{tt}}{2g_{tt}}&=0\notag\\
    C_3-\frac{\dot g_{rr}}{2g_{rr}}&=0\notag\\
    C_4+C_2\frac{g_{tt}}{g_{rr}}&=0.
\end{align}
We solve these equations in the obvious way for the functions $C_1$, $C_3$, and $C_4$. Now that we have explicit expression for these functions, we return to the flatness conditions and pick out the third equation. After plugging in the solutions to~\eqref{eq:ZeroNonMet} we find that it reduces to
\begin{equation}
    \dot C_2+\frac{C_2}{2}\left(\frac{\dot g_{tt}}{g_{tt}}-\frac{\dot g_{rr}}{g_{rr}}\right)=0,
\end{equation}
which is solved by
\begin{equation}
    C_2 = p\sqrt{\frac{g_{rr}}{g_{tt}}},
\end{equation}
where $p\in\mathbb{R}$ is an arbitrary integration constant. The flatness conditions now reduce to the two algebraic equations
\begin{align}
    c\,p&=0 & \text{and} && c^2 + p^2 - k &=0,
\end{align}
which have the four solutions
\begin{align}
    \label{eq:TCondition1} (c,p) &= (\pm\sqrt{k},0)\\
    \label{eq:TCondition2} (c,p) &=(0, \pm\sqrt{k}).
\end{align}
In the spatially flat case, $k=0$, these solutions further reduce to the single solution
\begin{equation}
    c=p=0.
\end{equation}
We call the connections obtained from the solution~\eqref{eq:TCondition1} $\Gamma_\T^{(\text{I}\pm)}$. It is explicitly given by
\begin{align}
    \Gamma^t{}_{\mu\nu}&=\left(
\begin{array}{cccc}
 \frac{\dot g_{tt}}{2 g_{tt}} & 0 & 0 & 0 \\
 0 & 0 & 0 & 0 \\
 0 & 0 & 0 & 0 \\
 0 & 0 & 0 & 0 \\
\end{array}
\right), &
\Gamma^r{}_{\mu\nu}&=\left(
\begin{array}{cccc}
 0 & \frac{\dot g_{rr}}{2 g_{rr}} & 0 & 0 \\
 0 & \frac{k\, r}{\chi} & 0 & 0 \\
 0 & 0 & -r \,\chi & \pm\sqrt{k}\, r^2 \, \chi\, \sin\theta \\
 0 & 0 & \mp\sqrt{k}\, r^2 \, \chi \,\sin\theta & -r\,\chi\,  \sin^2\theta \\
\end{array}
\right)\notag\\
\Gamma^{\theta}{}_{\mu\nu}&=\left(
\begin{array}{cccc}
 0 & 0 & \frac{\dot g_{rr}}{2 g_{rr}} & 0 \\
 0 & 0 & \frac{1}{r} & \mp \frac{\sqrt{k} \,\sin\theta}{\chi} \\
 0 & \frac{1}{r} & 0 & 0 \\
 0 & \pm \frac{\sqrt{k} \,\sin\theta}{\chi} & 0 & -\cos\theta\, \sin\theta \\
\end{array}
\right), & \Gamma^{\phi}{}_{\mu\nu}&=\left(
\begin{array}{cccc}
 0 & 0 & 0 & \frac{\dot g_{rr}}{2 g_{rr}} \\
 0 & 0 & \pm \frac{\sqrt{k} \,\csc\theta}{\chi} & \frac{1}{r} \\
 0 & \mp \frac{\sqrt{k} \,\csc\theta}{\chi} & 0 & \cot\theta \\
 0 & \frac{1}{r} & \cot\theta & 0 \\
\end{array}
\right).
\end{align}
Similarly, we denote the connection obtained from the solution~\eqref{eq:TCondition2} $\Gamma_\T^{(\text{II}\pm)}$. This connection is given by
\begin{align}
    \Gamma^t{}_{\mu\nu}&=\left(
\begin{array}{cccc}
 \frac{\dot g_{tt}}{2g_{tt}} & 0 & 0 & 0 \\
 0 & \pm \frac{1}{\chi}\sqrt{k\frac{g_{tt}}{g_{rr}}} & 0 & 0 \\
 0 & 0 & \pm r^2\sqrt{k\frac{g_{tt}}{g_{rr}}} & 0 \\
 0 & 0 & 0 & \pm r^2\,\sqrt{k\frac{g_{tt}}{g_{rr}}}\sin^2\theta \\
\end{array}
\right), & \Gamma^r{}_{\mu\nu}&=\left(
\begin{array}{cccc}
 0 & \frac{\dot g_{rr}}{2g_{rr}} & 0 & 0 \\
 \mp \sqrt{k\frac{g_{tt}}{g_{rr}}} & \frac{k\, r}{\chi} & 0 & 0 \\
 0 & 0 & -r\,\chi  & 0 \\
 0 & 0 & 0 & -r\,\chi  \sin^2\theta \\
\end{array}
\right)\notag\\
\Gamma^{\theta}{}_{\mu\nu}&=\left(
\begin{array}{cccc}
 0 & 0 & \frac{\dot g_{rr}}{2g_{rr}} & 0 \\
 0 & 0 & \frac{1}{r} & 0 \\
 \mp \sqrt{k\frac{g_{tt}}{g_{rr}}} & \frac{1}{r} & 0 & 0 \\
 0 & 0 & 0 & -\cos\theta\, \sin\theta \\
\end{array}
\right), & \Gamma^{\phi}{}_{\mu\nu}&=\left(
\begin{array}{cccc}
 0 & 0 & 0 & \frac{\dot g_{rr}}{2g_{rr}} \\
 0 & 0 & 0 & \frac{1}{r} \\
 0 & 0 & 0 & \cot\theta \\
 \mp \sqrt{k\frac{g_{tt}}{g_{rr}}} & \frac{1}{r} & \cot\theta & 0 \\
\end{array}
\right).
\end{align}
As noted above, when $k=0$ we obtain a single solution to the flatness conditions and we call the resulting connection~$\Gamma_\T^0$. It is explicitly given by
\begin{align}
    \Gamma^t{}_{\mu\nu}&=\left(
\begin{array}{cccc}
 \frac{\dot g_{tt}}{2g_{tt}} & 0 & 0 & 0 \\
 0 & 0 & 0 & 0 \\
 0 & 0 & 0 & 0 \\
 0 & 0 & 0 & 0 \\
\end{array}
\right), &
\Gamma^r{}_{\mu\nu}&=\left(
\begin{array}{cccc}
 0 &  \frac{\dot g_{rr}}{2g_{rr}} & 0 & 0 \\
 0 & 0 & 0 & 0 \\
 0 & 0 & -r & 0 \\
 0 & 0 & 0 & -r \sin^2\theta \\
\end{array}
\right)\notag\\
\Gamma^{\theta}{}_{\mu\nu}&=\left(
\begin{array}{cccc}
 0 & 0 & \frac{\dot g_{rr}}{2g_{rr}} & 0 \\
 0 & 0 & \frac{1}{r} & 0 \\
 0 & \frac{1}{r} & 0 & 0 \\
 0 & 0 & 0 & -\cos\theta \, \sin\theta \\
\end{array}
\right), & \Gamma^{\phi}{}_{\mu\nu}&=\left(
\begin{array}{cccc}
 0 & 0 & 0 & \frac{\dot g_{rr}}{2g_{rr}} \\
 0 & 0 & 0 & \frac{1}{r} \\
 0 & 0 & 0 & \cot\theta \\
 0 & \frac{1}{r} & \cot\theta & 0 \\
\end{array}
\right).
\end{align}
We observe that the connection is completely fixed in terms of the metric. There are no free functions left, apart from the metric components $g_{tt}$ and $g_{rr}$, which are as of yet undetermined. The connection for $f(\T)$ cosmology is therefore, in this sense, not dynamical. In other words: It does not propagate its own degrees of freedom.\\
\\
We also point out that the connection $\Gamma_\T^{0}$ can be obtained from the connections $\Gamma_\T^{(\text{I}\pm)}$ and $\Gamma_\T^{(\text{II}\pm)}$ by taking the limit $k\to 0$.\\
\\
So far we have not specified the sign of $k$ nor the signature of the metric. Our convention is such that $g_{tt}<0$ while $g_{rr}>0$. One can easily see that in the connections $\Gamma_\T^{(\text{I}\pm)}$ and $\Gamma_\T^{(\text{II}\pm)}$ this might lead to several imaginary components coming from $\sqrt{k}$ or $\sqrt{g_{tt}}$. However, the connection $\Gamma_\T^{(\text{I}\pm)}$ is real for $k>0$, while $\Gamma_\T^{(\text{II}\pm)}$ is real for $k<0$, see \cite{Hohmann:2019nat}, making $\Gamma_\T^{(\text{I}\pm)}$ applicable to spherical sections while $\Gamma_\T^{(\text{II}\pm)}$ can be used for hyperbolic ones.\\\\
\\
Before concluding this subsection and turning to the $f(\Q)$ connection we make a final observation: In the  MT literature, for instance in~\cite{Hohmann:2019}, one often uses the tetrad $e_\mu^a$ instead of the connection $\Gamma^\alpha{}_{\mu\nu}$. The former is defined by the equation
\begin{equation}
    g_{\mu\nu}=e^a_\mu e^b_\nu \eta_{ab},
\end{equation}
where $\eta_{ab}$ is the Minkowski metric. Using the tetrad, one can compute the torsion tensor as 
\begin{equation}
    T^\alpha{}_{\mu\nu}=e^\alpha_a \partial_\nu e^a_\mu.
\end{equation}
A popular choice in the literature (cf. for instance~\cite{Bahamonde:2021}) for the flat $k=0$ tetrad is simply the diagonal one; $e_\mu^a=\text{diag}(\sqrt{g_{tt}},\sqrt{g_{rr}},\sqrt{g_{rr}},\sqrt{g_{rr}})$. One can check that this tetrad is equivalent to the connection $\Gamma_\T^0$ we found in this subsection.

\subsection{Implementing Torsion-Freeness for the $f(\Q)$ Connection}\label{ssec:ZeroTorsion}
In order to implement the condition $T^\alpha{}_{\mu\nu} \overset{!}{=} 0$ we use the connection~\eqref{eq:SymmetricConnection} together with $C_5=c$. This leads to the simple conditions
\begin{align}
    C_3-C_4&=0\notag\\
    c&=0.
\end{align}
We thus set $c=0$ and replace $C_4$ by $C_3$. The flatness conditions then reduce to the three equations
\begin{align}
    \label{eq:RiemannQ1} C_1C_3-C_3^2-\dot C_3&=0\\
    \label{eq:RiemannQ2} C_1C_2-C_2C_3- \dot C_2 &=0\\
    \label{eq:RiemannQ33} k+C_2C_3&=0.
\end{align}
The last equation is of particular interest. In the spatially flat case, $k=0$, it implies that either $C_2 = 0$, or $C_3 = 0$, or both vanish. For $k\neq 0$ we must have\footnote{We do not consider cases where one of the two functions vanishes while the other one diverges while $C_2 C_3 = -k$ is kept fixed.} both $C_2$ and $C_3$ non-vanishing to have $C_2C_3=-k\neq 0$. We thus have to discuss the spatially flat case separately. For now, we keep $k\neq 0$. We can then fix $C_3=-k/C_2$, which reduces the remaining flatness conditions to the single equation
\begin{equation}
    k+C_1C_2+\dot C_2=0.
\end{equation}
Since $C_2\neq 0$, we can solve the above equation for $C_1$ and we obtain
\begin{equation}
    C_1=-\frac{k+\dot C_2}{C_2}.
\end{equation}
With this we have solved all conditions and we call the resulting connection $\Gamma^{(k)}_\Q$. It is explicitly given by
\begin{align}
    \Gamma^t{}_{\mu\nu}&=\left(
\begin{array}{cccc}
 -\frac{k+\dot C_2}{C_2} & 0 & 0 & 0 \\
 0 & \frac{C_2}{\chi} & 0 & 0 \\
 0 & 0 & r^2\, C_2 & 0 \\
 0 & 0 & 0 & r^2 \sin^2\theta\, C_2 \\
\end{array}
\right), & \Gamma^r{}_{\mu\nu}&= \left(
\begin{array}{cccc}
 0 & -\frac{k}{C_2} & 0 & 0 \\
 -\frac{k}{C_2} & \frac{k\, r}{\chi} & 0 & 0 \\
 0 & 0 & -r\,\chi  & 0 \\
 0 & 0 & 0 & -r\,\chi  \sin^2\theta \\
\end{array}
\right)\notag\\
\Gamma^{\theta}{}_{\mu\nu}&=\left(
\begin{array}{cccc}
 0 & 0 & -\frac{k}{C_2} & 0 \\
 0 & 0 & \frac{1}{r} & 0 \\
 -\frac{k}{C_2} & \frac{1}{r} & 0 & 0 \\
 0 & 0 & 0 & -\cos\theta\, \sin\theta \\
\end{array}
\right), & \Gamma^{\phi}{}_{\mu\nu}& = \left(
\begin{array}{cccc}
 0 & 0 & 0 & -\frac{k}{C_2} \\
 0 & 0 & 0 & \frac{1}{r} \\
 0 & 0 & 0 & \cot\theta \\
 -\frac{k}{C_2} & \frac{1}{r} & \cot\theta & 0 \\
\end{array}
\right).
\end{align}
Notice that the connection still contains a free function, $C_2(t)$, which is potentially a connection degree of freedom that can influence the metric.\\
\\
Let us now turn to the spatially flat case described by $k=0$. We then have three possible solutions of~\eqref{eq:RiemannQ33}, namely $C_2=0$, $C_3\neq 0$, which we call case (I), $C_2\neq 0$, $C_3=0$ called case (II), and $C_2=C_3=0$ called case (III).\\
\\
In case (I) the condition~\eqref{eq:RiemannQ2} is trivially fulfilled, and condition~\eqref{eq:RiemannQ1} can be solved for $C_1$ since $C_3\neq 0$:
\begin{equation}
    C_1=C_3+\frac{\dot C_3}{C_3}.
\end{equation}
The connection which results from choosing case (I) is denoted by  $\Gamma_\Q^\text{(I)}$ and it is explicitly given by
\begin{align}
\Gamma^t{}_{\mu\nu}&=\left(
\begin{array}{cccc}
 C_3+\frac{\dot C_3}{C_3} & 0 & 0 & 0 \\
 0 & 0 & 0 & 0 \\
 0 & 0 & 0 & 0 \\
 0 & 0 & 0 & 0 \\
\end{array}
\right), &
\Gamma^r{}_{\mu\nu}&=\left(
\begin{array}{cccc}
 0 & C_3 & 0 & 0 \\
 C_3 & 0 & 0 & 0 \\
 0 & 0 & -r & 0 \\
 0 & 0 & 0 & -r \sin^2\theta \\
\end{array}
\right)\notag\\
\Gamma^{\theta}{}_{\mu\nu}&=\left(
\begin{array}{cccc}
 0 & 0 & C_3 & 0 \\
 0 & 0 & \frac{1}{r} & 0 \\
 C_3 & \frac{1}{r} & 0 & 0 \\
 0 & 0 & 0 & -\cos \theta \, \sin\theta \\
\end{array}
\right), &
\Gamma^{\phi}{}_{\mu\nu}&=\left(
\begin{array}{cccc}
 0 & 0 & 0 & C_3 \\
 0 & 0 & 0 & \frac{1}{r} \\
 0 & 0 & 0 & \cot\theta  \\
 C_3 & \frac{1}{r} & \cot\theta & 0 \\
\end{array}
\right).
\end{align}
The function $C_3(t)$ is unspecified and explicitly appears in the connection $\Gamma_\Q^\text{(I)}$. It therefore constitutes a potential connection degree of freedom. For the second case, i.e., case (II), one finds that the condition~\eqref{eq:RiemannQ1} is trivially fulfilled while condition~\eqref{eq:RiemannQ2} can be solved for $C_1$ because $C_2\neq 0$. One obtains
\begin{equation}
    C_1=-\frac{\dot C_2}{C_2}.
\end{equation}
We denote the connection which results from case (II) by $\Gamma_\Q^\text{(II)}$. It is explicitly given by
\begin{align}
    \Gamma^t{}_{\mu\nu}&=\left(
\begin{array}{cccc}
 -\frac{\dot C_2}{C_2} & 0 & 0 & 0 \\
 0 & C_2 & 0 & 0 \\
 0 & 0 & r^2\, C_2 & 0 \\
 0 & 0 & 0 & r^2 \sin^2\theta\, C_2 \\
\end{array}
\right), &
\Gamma^r{}_{\mu\nu}& = \left(
\begin{array}{cccc}
 0 & 0 & 0 & 0 \\
 0 & 0 & 0 & 0 \\
 0 & 0 & -r & 0 \\
 0 & 0 & 0 & -r \sin^2\theta \\
\end{array}
\right)\notag\\
\Gamma^{\theta}{}_{\mu\nu}& = \left(
\begin{array}{cccc}
 0 & 0 & 0 & 0 \\
 0 & 0 & \frac{1}{r} & 0 \\
 0 & \frac{1}{r} & 0 & 0 \\
 0 & 0 & 0 & -\cos\theta\, \sin\theta\\
\end{array}
\right), & \Gamma^{\phi}{}_{\mu\nu}& = \left(
\begin{array}{cccc}
 0 & 0 & 0 & 0 \\
 0 & 0 & 0 & \frac{1}{r} \\
 0 & 0 & 0 & \cot\theta \\
 0 & \frac{1}{r} & \cot\theta & 0 \\
\end{array}
\right).
\end{align}
In this connection, the function $C_2(t)$ explicitly appears and it is left unspecified. Hence, we find again that the connection possesses a potential degree of freedom which it can propagate and which could influence the metric. Finally, we turn to case (III). Now all flatness conditions are trivially fulfilled. We call the connection which results from case (III) $\Gamma_\Q^\text{(III)}$ and its form is given by
\begin{align}
    \Gamma^t{}_{\mu\nu}& = \left(
\begin{array}{cccc}
 C_1 & 0 & 0 & 0 \\
 0 & 0 & 0 & 0 \\
 0 & 0 & 0 & 0 \\
 0 & 0 & 0 & 0 \\
\end{array}
\right), & 
\Gamma^r{}_{\mu\nu}& = \left(
\begin{array}{cccc}
 0 & 0 & 0 & 0 \\
 0 & 0 & 0 & 0 \\
 0 & 0 & -r & 0 \\
 0 & 0 & 0 & -r \sin^2\theta \\
\end{array}
\right)\notag\\
\Gamma^{\theta}{}_{\mu\nu}&=\left(
\begin{array}{cccc}
 0 & 0 & 0 & 0 \\
 0 & 0 & \frac{1}{r} & 0 \\
 0 & \frac{1}{r} & 0 & 0 \\
 0 & 0 & 0 & -\cos\theta\, \sin\theta \\
\end{array}
\right),&
\Gamma^{\phi}{}_{\mu\nu}&=\left(
\begin{array}{cccc}
 0 & 0 & 0 & 0 \\
 0 & 0 & 0 & \frac{1}{r} \\
 0 & 0 & 0 & \cot\theta \\
 0 & \frac{1}{r} & \cot\theta & 0 \\
\end{array}
\right).
\end{align}
Also in this case do we find an unspecified function which explicitly appears in the connection -- the function $C_1(t)$. Hence, all three connections, $\Gamma_\Q^\text{(I)}$, $\Gamma_\Q^\text{(II)}$, and $\Gamma_\Q^\text{(III)}$, contain a potential degree of freedom. Whether or not there actually is a propagating degree of freedom stemming from the connection can ultimately only be answered by studying the field equations. This will be the subject of section~\ref{sec:FieldEquations}.\\
\\
Before concluding this section and moving on to study the symmetry reduced field equations, we observe some interesting limits. In fact, the connection  $\Gamma_\Q^\text{(III)}$ can be obtained as a limiting case from the connections $\Gamma_\Q^\text{(I)}$ and $\Gamma_\Q^\text{(II)}$. To obtain $\Gamma_\Q^\text{(III)}$ from $\Gamma_\Q^\text{(I)}$ one has to set $C_3=\lambda\Phi(t)$, with arbitrary $\Phi(t)$, and let $\lambda\to 0$. This leads to an arbitrary $C_1=\dot\Phi/\Phi$.\\
In order to obtain the connection  $\Gamma_\Q^\text{(III)}$ from  $\Gamma_\Q^\text{(II)}$ one has to set $C_2=\lambda\Phi(t)$, with arbitrary $\Phi(t)$, and let $\lambda\to 0$. This consequently leads to an  arbitrary $C_1=-\dot\Phi/\Phi$.\\
\\
Furthermore, one  can obtain the $k=0$ cases as limits from the connection $\Gamma_\Q^{(k)}$. This is achieved by taking  $k\to 0$ while keeping $C_2\neq 0$. This produces the connection $\Gamma_\Q^\text{(II)}$. In turn, replacing first $C_2=-k/C_3$ in $\Gamma_\Q^{(k)}$ with $C_3\neq 0$ and then taking $k\to 0$, one obtains the connection $\Gamma_\Q^\text{(I)}$. We can also replace $C_2=\sqrt{|k|}\,\Phi$, with arbitrary $\Phi$, and then take $k\to 0$. This leads  to the connection $\Gamma_\Q^\text{(I)}$ with arbitrary $C_1=-\dot\Phi/\Phi$.\\
\\
Finally, let us point out that the connections which have been reported in this section completely agree with the connections which have recently appeared in~\cite{Hohmann:2021} and which were found using the same approach. 

\section{Cosmological Field Equations}\label{sec:FieldEquations}\setcounter{equation}{0}
In section~\ref{sec:SymmReduction} we have performed a symmetry-reduction of the metric and of the connection. This gave us the metric~\eqref{eq:SymmetricMetric} and the connection~\eqref{eq:SymmetricConnection}, which both respect the cosmological principles of homogeneity and isotropy. In the following section we have then separately implemented the geometric postulates of $f(\T)$ and $f(\Q)$ gravity. This gave rise to the connections $\Gamma_\Q^{(\text{I})}$, $\Gamma_\Q^{\text{II}}$, $\Gamma^{(\text{III})}_\Q$, and $\Gamma^{(k)}_\Q$ for $f(\Q)$ cosmology and  $\Gamma_\T^{(0)}$, $\Gamma^{(\text{I}_\pm)}_\T$, and $\Gamma^{(\text{II}_\pm)}_\T$ for $f(\T)$~cosmology.

We are now therefore fully equipped to study the symmetry reduced field equations which describe $f(\T)$ and $f(\Q)$ cosmology. We first study the equations for $f(\T)$ and then turn our attention to $f(\Q)$ cosmology.
\subsection{$f(\T)$ Cosmology}\label{ssec:f(T)}
In order to obtain the field equations for $f(\T)$ cosmology we simply plug in the metric~\eqref{eq:SymmetricMetric} and the various connections $\Gamma_\T^{(0,\text{I}\pm,\text{II}\pm)}$ into the field equations~\eqref{eq:MTFieldEqs}. As we have seen in the previous section, the connections $\Gamma_\T^{(0,\text{I}\pm,\text{II}\pm)}$ do not depend on any free functions. They are completely determined by the metric. Hence, we expect that the connection is not dynamical and indeed we find that the connection field equations are all identically satisfied; $\mathcal{C}_{\alpha\beta}=0$. For the metric field equations we find that they all vanish, except for $\M^t{}_t,$ and $\M^r{}_r=\M^\theta{}_\theta=\M^\phi{}_\phi$.\\
In the sequel we use the standard convention that $g_{tt}=-\N(t)$, where $\N$ is the lapse function, and we set $g_{rr}=a(t)^2$, where $a$ is the scale factor. We also introduce the Hubble  parameter  $H:=\dot a/a$. Now we can discuss the field equations for each connection $\Gamma_\T^{(0,\text{I}\pm,\text{II}\pm)}$  individually:\\
\\
For the connection choice $\Gamma_\T^{0}$, which corresponds to spatially flat sections, $k=0$, one finds the equations
\begin{align}
	\T &= -\frac{6 H^2}{\mathcal{N}}\notag\\
    \M^t{}_{t} &= -\frac{6 H^2 f'}{\mathcal{N}}-\frac{f}{2}, & \M^r{}_{r} &= -\frac{2 f' \left(2 \mathcal{N} \dot H+H \left(6 H \mathcal{N}-\dot{\mathcal{N}}\right)\right)+\mathcal{N} \left(4 H f'' \dot{\mathbb{T}}+f \mathcal{N}\right)}{2 \mathcal{N}^2} \label{eq:EoMT0}.
\end{align}
These are the field equations found in the literature for the diagonal tetrad \cite{Bahamonde:2021}.\\
\\
For the choice $\Gamma_\T^{(\text{I}\pm)}$ one finds
\begin{align}\label{eq:EQsGIpm}
	\T &= \frac{6 k}{a^2}-\frac{6 H^2}{\mathcal{N}}\notag\\
    \M^t{}_{t} &= -\frac{6 H^2 f'}{\mathcal{N}}-\frac{f}{2},&
    \M^r{}_{r} &= \frac{2 f' \left(2 \mathcal{N} \left(\frac{k \mathcal{N}}{a^2}-3 H^2\right)-2 \mathcal{N} \dot H+H \dot{\mathcal{N}}\right)-\mathcal{N} \left(4 H f'' \dot{\mathbb{T}}+f \mathcal{N}\right)}{2 \mathcal{N}^2}.
\end{align}
Notice that the field equations~\eqref{eq:EQsGIpm} are the same for either choice of sign because only $k$, rather than $\pm\sqrt{k}$, enters in these equations. The field equations are also manifestly real. This is not always the case for the remaining connection choice, i.e., for $\Gamma_\T^{(\text{II}\pm)}$. For this last case one finds the field equations
\begin{align}
	\T &= -\frac{6 \left(a^2 \pm a\sqrt{k} \sqrt{-\mathcal{N}}\right)^2}{a^4 \mathcal{N}}\notag \\
    \M^r{}_{r} &= \frac{2 f' \left(-2 a^2 \mathcal{N} \dot H-6 a^2 H^2 \mathcal{N}\pm 6 a H \sqrt{k} (-\mathcal{N})^{3/2}+a^2 H \dot{\mathcal{N}}+2 k \mathcal{N}^2\right)-\mathcal{N} \left(4 f'' \dot{\T} \left(a^2 H\pm a \sqrt{k} \sqrt{-\mathcal{N}}\right)+a^2 f \mathcal{N}\right)}{2 a^2 \mathcal{N}^2}\notag\\
    \M^t{}_{t} &= -\frac{6 H f' \left(\frac{\pm\sqrt{k} \sqrt{-\mathcal{N}}}{a}+H\right)}{\mathcal{N}}-\frac{f}{2}.
\end{align}
Notice that in this case the factor $\pm\sqrt{k}$ explicitly enters the equations of motion and hence we get different equations, depending on which sign we choose for $\Gamma_\T^{(\text{II}\pm)}$. Moreover, there are terms of the form $\sqrt{-\N}$. These equations are thus  complex --and hence not viable-- for $k>0$ and $\mathcal N>0$.

\subsection{$f(\Q)$ Cosmology}\label{ssec:f(Q)}
To perform the symmetry-reduction of the $f(\Q)$ field equations we use again the metric~\eqref{eq:SymmetricMetric} and the different connections~$\Gamma_\Q^{(\text{I,II,III},k)}$ we found in the previous section. For some connection choices we find indeed that the connection is dynamical, because the arbitrary functions appearing in the connection come along with their own equations of motion; $\mathcal{C}_t$. All other connection field equations $\mathcal{C}_\alpha$ are identically zero. Moreover, all metric field equations are identically satisfied with the exception of  $\M^t{}_t$ and $\M^r{}_r=\M^\theta{}_\theta=\M^\phi{}_\phi$. Below we report these field equations for each choice of connection separately. As in the case of $f(\T)$ cosmology, we write these equations in terms of lapse $\N$ and scale factor $a$.\\
\\
We begin with the connection choice  $\Gamma_\Q^{(\text{I})}$ and find
\begin{align}\label{eq:EoMGamma1Q}
    \Q &= -\frac{3 \left(C_3 \left(\dot{\mathcal{N}}-6 H \mathcal{N}\right)-2 \mathcal{N} \dot C_3+4 H^2 \mathcal{N}\right)}{2 \mathcal{N}^2}\notag\\
    \M^t{}_{t} &= -\frac{3 f' \left(C_3 \left(\dot{\mathcal{N}}-6 H \mathcal{N}\right)-2 \mathcal{N} \dot C_3+8 H^2 \mathcal{N}\right)+2 \mathcal{N} \left(3 C_3 f'' \dot{\mathbb{Q}}+f \mathcal{N}\right)}{4 \mathcal{N}^2}\notag\\
    \M^r{}_{r} &= \frac{f' \left(3 C_3 \left(6 H \mathcal{N}-\dot{\mathcal{N}}\right)+6 \mathcal{N} \dot{C}_3-8 \mathcal{N} \dot{H}-24 H^2 \mathcal{N}+4 H \dot{\mathcal{N}}\right)-2 \mathcal{N} \left(\left(4 H-3 C_3\right) f'' \dot{\mathbb{Q}}+f \mathcal{N}\right)}{4 \mathcal{N}^2}\notag\\
    \C_t &= -\frac{3 C_3 \left(f'' \dot{\mathbb{Q}} \left(6 H \mathcal{N}-\dot{\mathcal{N}}\right)+2 f^{(3)} \mathcal{N} \dot{\mathbb{Q}}^2+2 \mathcal{N} f'' \ddot{\mathbb{Q}}\right)}{4 \mathcal{N}^2}.
\end{align}
The connection component $C_3$ is dynamical with equation of motion $\C_t$ and can influence the metric though $\M^t{}_{t}$ and $\M^r{}_{r}$.\\ 
For the choice $\Gamma_\Q^{(\text{II})}$ we find instead 
\begin{align}\label{eq:EoMGamma2Q}
    \Q &= \frac{3 \left(C_2 \left(6 H \mathcal{N}+\dot{\mathcal{N}}\right)+2 \mathcal{N} \dot C_2-4 H^2\right)}{2 \mathcal{N}}\notag\\
    \M^t{}_{t} &= \frac{3 f' \left(C_2 \left(6 H \mathcal{N}+\dot{\mathcal{N}}\right)+2 \mathcal{N} \dot C_2-8 H^2\right)-2 \mathcal{N} \left(f-3 C_2 f'' \dot{\mathbb{Q}}\right)}{4 \mathcal{N}}\notag\\
    \M^r{}_{r} &= \frac{f' \left(3 C_2 \mathcal{N} \left(6 H \mathcal{N}+\dot{\mathcal{N}}\right)+6 \mathcal{N}^2 \dot C_2-8 \mathcal{N} \dot H-24 H^2 \mathcal{N}+4 H \dot{\mathcal{N}}\right)-2 \mathcal{N} \left(f'' \dot{\mathbb{Q}}\left(4 H-C_2 \mathcal{N}\right)+f \mathcal{N}\right)}{4 \mathcal{N}^2}\notag\\
    \C_t &= \frac{3 C_2 \left(f'' \dot{\mathbb{Q}} \left(10 H \mathcal{N}+\dot{\mathcal{N}}\right)+2 f^{(3)} \mathcal{N} \dot{\mathbb{Q}}^2+2 \mathcal{N} f'' \ddot{\mathbb{Q}}\right)}{4 \mathcal{N}}+3 \dot C_2 f'' \dot{\mathbb{Q}}. 
\end{align}
Now the connection component $C_2$ is dynamical with equation of motion $\C_t$ and it can influence the metric though $\M^t{}_{t}$ and $\M^r{}_{r}$. The situation is quite different for the choice  $\Gamma_\Q^{(\text{III})}$ for which one finds 
\begin{align}\label{eq:EoMGamma3Q}
    \Q &= -\frac{6 H^2}{\mathcal{N}}\notag\\
    \M^t{}_{t} &= -\frac{6 H^2 f'}{\mathcal{N}}-\frac{f}{2}\notag\\
    \M^r{}_{r} &= -\frac{6 H^2 f'}{\mathcal{N}}-\frac{f}{2}\notag\\
    \C_t &= 0. 
\end{align}
In particular, the function $C_1$ does not appear in these equations at all and  $\mathcal{C}_t$ is trivially satisfied. In this case, the connection is not dynamical and the function $C_1$ remains undetermined by the equations of motion. This case is especially interesting, as the connection is then effectively given --by setting $C_1=0$-- by the spherical connection $\Gamma^s$ corresponding to the coincident gauge. These equations of motion were previously reported in~\cite{BeltranJimenez:2017,Jimenez:2018,Mandal:2020}. Finally, we can also choose $\Gamma_\Q^{(k)}$ for non-flat spatial sections $k\neq 0$  and we find
\begin{align}\label{eq:EoMGammakQ}
    \Q &= \frac{3 \left(2 C_2^2 \mathcal{N} \left(\frac{2 k \mathcal{N}}{a^2}+\mathcal{N} \dot C_2-2 H^2\right)+\frac{C_2 k \left(\dot{\mathcal{N}}-2 H \mathcal{N}\right)}{a^2}+\frac{2 k \mathcal{N} \dot C_2}{a^2}+C_2^3 \mathcal{N} \left(6 H \mathcal{N}+\dot{\mathcal{N}}\right)\right)}{2 C_2^2 \mathcal{N}^2}\notag\\
    \M^t{}_{t} &= \frac{\frac{3 C_2 k \left(f' \left(\dot{\mathcal{N}}-2 H \mathcal{N}\right)+2 \mathcal{N} f'' \dot{\mathbb{Q}}\right)}{a^2}+\frac{6 k \mathcal{N} \dot C_2 f'}{a^2}-2 C_2^2 \mathcal{N} \left(3 f' \left(4 H^2-\mathcal{N} \dot C_2\right)+f \mathcal{N}\right)+3 C_2^3 \mathcal{N} \left(f' \left(6 H \mathcal{N}+\dot{\mathcal{N}}\right)+2 \mathcal{N} f'' \dot{\mathbb{Q}}\right)}{4 C_2^2 \mathcal{N}^2}\notag\\
    \M^r{}_{r} &= \frac{1}{4 C_2^2 \mathcal{N}^2}\Big[ \frac{3 C_2 k \left(f' \left(\dot{\mathcal{N}}-2 H \mathcal{N}\right)-2 \mathcal{N} f'' \dot{\mathbb{Q}}\right)}{a^2}+\frac{6 k \mathcal{N} \dot C_2 f'}{a^2}+C_2^3 \mathcal{N} \left(3 f' \left(6 H \mathcal{N}+\dot{\mathcal{N}}\right)+2 \mathcal{N} f'' \dot{\mathbb{Q}}\right)-\notag\\
    &-2 C_2^2 \left(f' \left(\mathcal{N} \left(-\frac{4 k \mathcal{N}}{a^2}-3 \mathcal{N} \dot C_2+12 H^2\right)+4 \mathcal{N} \dot H-2 H \dot{\mathcal{N}}\right)+\mathcal{N} \left(4 H f'' \dot{\mathbb{Q}}+f \mathcal{N}\right)\right)\Big]\notag\\
    \mathcal{C}_t &= \frac{3 f'' \dot{\mathbb{Q}} \left(\frac{k \left(6 H \mathcal{N}-\dot{\mathcal{N}}\right)}{a^2}+C_2^2 \mathcal{N} \left(10 H \mathcal{N}+\dot{\mathcal{N}}\right)+4 C_2 \mathcal{N}^2 \dot C_2\right)+6 f^{(3)} \mathcal{N} \dot{\mathbb{Q}}^2 \left(\frac{k}{a^2}+C_2^2 \mathcal{N}\right)+6 \mathcal{N} f'' \ddot{\mathbb{Q}} \left(\frac{k}{a^2}+C_2^2 \mathcal{N}\right)}{4 C_2 \mathcal{N}^2}.
\end{align}
The scale factor now appears explicitly only in terms of the form $k/a^2$.

\subsection{Comparison of $f(\T)$ and $f(\Q)$ Cosmologies}\label{ssec:Comparison}
The largest difference between $f(\T)$ and $f(\Q)$ cosmology can be found in the difference of admissible connection choices. While the $f(\T)$ connections are completely fixed in terms of the metric, the $f(\Q)$ connections admit freely specifiable functions $C_i(t)$. For instance, compare~\eqref{eq:EoMT0} with~\eqref{eq:EoMGamma1Q} or with~\eqref{eq:EoMGamma2Q} for the spatially flat case, $k=0$. In some cases, as discussed in the previous subsection, these free functions are promoted to true connection degrees of freedom. They do not only enter the metric field equations, they possess their own equations of motion and they genuinely influence the gravitational field. It is therefore fair to say that $f(\Q)$ cosmology is much richer than $f(\T)$ cosmology at the background level because of these additional degrees of freedom in the form of $C_i$ functions. This is true for flat as well as for non-flat spatial sections. Some solutions to the $f(\Q)$ field equations exploiting the connection degrees of freedom will be discussed in the next section.\\
A possible explanation is that while both connections in $f(\T)$ and $f(\Q)$ gravity respect homogeneity and isotropy and are flat, the $f(\T)$ connection has also to satisfy the conditions $Q_{\alpha\mu\nu}=0$. These are 40 independent equations, as opposed to the 24 independent equations the $f(\Q)$ connection has to satisfy because of the conditions $T^\alpha{}_{\mu\nu}=0$. Hence, it seems that the $f(\T)$ connection is in general more restricted than the $f(\Q)$ connection. A similar feature was found for stationary and spherically symmetric spacetimes~\cite{DAmbrosio:2021}.\\
However, there are also some similarities between $f(\T)$ and $f(\Q)$ gravity. In particular, in the flat case\footnote{We have not found an analogous identity for $k\neq 0$.}, where $k=0$, we see that the field equations of $f(\T)$ cosmology agree exactly with those of $f(\Q)$ cosmology expressed in  coincident gauge with connection $\Gamma^s$, i.e. \eqref{eq:EoMT0} $=$ \eqref{eq:EoMGamma3Q}. In particular, we find $\T=\Q$. Thus, for some connection choices, $f(\T)$ and $f(\Q)$ gravity give the same cosmologies. However, for $k\neq 0$  one generally finds different field equations and therefore also expects different cosmologies for  $f(\T)$ and $f(\Q)$ gravity. 

\section{GR and beyond-GR Solutions}\label{sec:Examples}\setcounter{equation}{0}
We present two examples which illustrate how one can use the free connection variables in $f(\Q)$ cosmology to construct GR solutions for arbitrary $f$ as well as beyond-GR solutions for specific choices of $f$. 

\subsection{Exact GR Solutions in $f(\Q)$ Cosmology}\label{ssec:ExampleGR}
Let us choose the $\Gamma_\Q^k$ connection of $f(\Q)$ cosmology for any value of $k$. It is implicitly understood that $k=0$ is obtained by taking a limit. We then start our considerations by first looking at the non-metricity scalar $\Q$ given by~\eqref{eq:EoMGammakQ}. Since $\Q$ contains the function $C_2$, we can at least in principle solve the equation $\Q=\Q_0$, where $\Q_0\in\mathbb R$ is a constant, for $C_2$. For example, in the case $k=0$ one finds from equation~\eqref{eq:EoMGamma2Q}
\begin{equation}
    C_2=\frac{1}{a^3\sqrt{\N}}\left(l-\frac{1}{3}\int^t \frac{a(\tau)^3}{\sqrt{\N(\tau)}}\left(\Q_0\, \N(\tau)+6H(\tau)^2\right)\,\dd \tau\right),
\end{equation}
with arbitrary integration constant $l\in\mathbb R$. Because $\Q$ is now constant, one can easily check from the form~\eqref{eq:RewrittenMetricFieldEq} that we can write the metric field equations as
\begin{equation}\label{eq:GREquations}
	G_{\mu\nu} + \Lambda_\textsf{eff} \, g_{\mu\nu} = \bar{T}_{\mu\nu},
\end{equation}
where we have defined
\begin{align}
	\Lambda_\textsf{eff} &:= \frac{1}{2}\frac{f(\Q_0)-f'(\Q_0)\Q_0}{f'(\Q_0)}\notag\\
	\bar{T}_{\mu\nu} &:= \frac{1}{f'(\Q_0)} T_{\mu\nu}.
\end{align}
Moreover, the connection field equation $\C_t$, \eqref{eq:EoMGammakQ} or \eqref{eq:EoMGamma2Q}, is identically satisfied for $\Q=\Q_0$. We have thus shown that in $f(\Q)$ cosmology we can choose the connection such that for any $f$ we recover the exact GR solutions. This is analogous to the result derived in~\cite{DAmbrosio:2021} where it was shown that in $f(\Q)$ as well as in $f(\T)$ gravity one can recover the Schwarzschild-deSitter-Nordstr\"{o}m solution for any choice of $f$.\\
\\
We note in particular that in $f(\T)$ gravity an analogous procedure cannot be done, as the connection is completely fixed; the only free functions in $\T$ is the metric given by $\N$ and $a$. This is in contrast to what was found in~\cite{DAmbrosio:2021} for $f(\T)$ black holes, as for stationary and spherically symmetric solutions the $f(\T)$ connection is a priory --that means before using the equations of motion-- free, and one can use it to set $\T=\T_0$ to a constant, leading to exact GR solutions for any $f$ via~\eqref{eq:MTSimpleMetEq}.

\subsection{Exact connection-driven Vacuum Solutions for $f(\Q)=\Q^\kappa$}\label{ssec:ExampleExact}
Using a similar strategy as in~\cite{DAmbrosio:2021}, we can prove that $f(\Q)$ cosmology possesses exact beyond-GR solutions. The non-trivial dynamics of the scale factor are here driven by the connection degree of freedom. Hence, the procedure presented here will not work in $f(\T)$ cosmology. We choose the connection $\Gamma_\Q^\text{(II)}$ and set the lapse to unity, $\N=1$. We also make the choice\footnote{The following procedure does not work for perturbations from ST, i.e. $f(\Q)=\Q+F(\Q)$ for some function $F$, e.g. $F(\Q)=\alpha \Q^\kappa$.} $f(\Q)=\Q^\kappa$ with integer $\kappa\geq 2$.  This implies in particular that the solution will not be of the form ``GR plus corrections'', but will in general deviate significantly from GR.\\
\\
Under these assumptions and with these choices, the field equations are then given by $\M^t{}_{t}$, $\M^r{}_{r}$, and $\mathcal{C}_t$ as expressed in~\eqref{eq:EoMGamma2Q}. Curiously, one can check that in fact all three equations have a common multiplicative factor in front. Namely
\begin{equation}
    \M^t{}_{t},\,\M^r{}_{r},\, \mathcal{C}_t \propto \left(\dot C_2-2H^2+3HC_2\right)^{\kappa-2}.
\end{equation}
For $\kappa\geq 3$ we can then trivially solve the vacuum equations, i.e., assuming $\rho=p=0$, by setting
\begin{equation}\label{eq:ExactBeyondGR}
    \dot C_2-2H^2+3HC_2=0.
\end{equation}
But since $C_2$ can now be freely chosen, this does not restrict the scale factor at all. For any $a$ one can choose the corresponding $C_2$ as
\begin{equation}
    C_2=\frac{l}{a^3}+\frac{2}{a^3}\int^t a(\tau)\dot a(\tau)^2\,\dd \tau,
\end{equation}
with an arbitrary integration constant $l$. Alternatively, one can express for any given choice of $C_1$ the scale factor as
\begin{equation}\label{eq:BeyondGRa}
    a=a_0 \exp\left(\int^t\left( \frac{3C_2(\tau)}{4}\pm \sqrt{\frac{9}{16}C_2(\tau)^2+\frac{1}{2}\dot C_2(\tau)}\right)\dd \tau\right),
\end{equation}
where now $a_0$ is an arbitrary integration constant. As an example, let us choose
\begin{equation}
    C_2=-\frac{2\lambda^2}{1-3\lambda}\frac{1}{t}.
\end{equation}
We then find from equation~\eqref{eq:ExactBeyondGR}
\begin{equation}
    a=a_0\, t^\lambda.
\end{equation}
Hence, for $\lambda=\frac{2}{3(1+w)}$ we can mimic the scale factor of a fluid with equation of state $p=w\rho$, $w$ constant. For the choice
\begin{equation}
    C_2=\frac{2}{3}\sqrt{\frac{\Lambda}{3}},
\end{equation}
on the other hand, one finds deSitter space:
\begin{equation}\label{eq:C2DrivendS}
    a=a_0 \exp\left(\sqrt{\frac{\Lambda}{3}}t\right).
\end{equation}
Even though we are in the vacuum case, $\rho=p=0$, the connection can drive arbitrary solutions in $\Q^\kappa$ gravity with $\kappa\geq 3$. In GR, on the other hand, in the vacuum case the scale factor has to be constant,  $a=a_0$; this is in fact true for all flat $k=0$ $f(\Q)$ cosmologies that use the spherical connection $\Gamma^s$, as one can easily see that~\eqref{eq:EoMGamma3Q} with $\rho=p=0$ and $\N=1$ imply constant $H$. This is also captured by~\eqref{eq:BeyondGRa} as $a\to a_0$ for $C_2\to 0$, which implies $\Gamma_\Q^{(\text{II})}\to\Gamma^s$. The connection component $C_2$ can thus mimic arbitrary fluid matter leading to an arbitrary scale factor~$a$.\\
\\
This new solution could be of interest in the early universe, where $\rho,p\approx 0$. The connection degree of freedom $C_2$ could  drive inflation by means of~\eqref{eq:C2DrivendS}. However, for $\rho,p\neq 0$ the complicated equations of motion cannot be solved so easily and the solution for the non-vacuum case could depart significantly from~\eqref{eq:BeyondGRa}.

\section{Conclusion}\label{sec:Conclusion}\setcounter{equation}{0}
In this paper we have carried out a systematic symmetry-reduction of the metric and the connection in order to analyze $f(\T)$ and $f(\Q)$ cosmology. We have derived the most general metric and the most general connections which respect homogeneity and isotropy and, on top of that, which also satisfy the geometric postulates of Metric Teleparallelism and Symmetric Teleparallelism. In particular, we found that the $f(\T)$ connection is completely fixed in terms of the metric, while the  $f(\Q)$ connections contain free functions. Our analysis of the symmetry-reduced field equations has further revealed that some of these $f(\Q)$ connections become dynamical, i.e., they possess their own equations of motion and they will in general also influence the metric. 

Moreover, we observed that for spatially flat spacetimes, $k=0$, the spherical connection $\Gamma^s$ of $f(\Q)$ cosmology produces the same field equations as the ones in $f(\T)$ cosmology. In that particular case, the torsion scalar and the non-metricity scalar are equal to each other, $\T = \Q$. Hence, even though $f(\Q)$ cosmology has a richer structure than $f(\T)$ cosmology, there is some overlap between the two models.
\\
The situation can be compared with spherically symmetric and stationary spacetimes in $f(\T)$ and $f(\Q)$ gravity. The pertinent field equations were presented in~\cite{DAmbrosio:2021}, and the same situation was found as for cosmology. Namely, both $f(\T)$ and $f(\Q)$ gravity admitted beyond-GR solutions for various choices of the connections, but in $f(\T)$ the connection is fixed -- the connection field equations turned out to be mere constraints, rather than dynamical equations\footnote{In particular, the admissible $f(\T)$ connections initially contained undetermined functions, which could be used to construct constant $\T$ solutions --which are then exactly the GR solutions-- analogously to what we did here for $f(\Q)$ (see section~\ref{ssec:ExampleGR}). This procedure, as we recall, does not work for $f(\T)$ cosmology. But it can be used to construct GR solutions for $f(\Q)$ black hole spacetimes~\cite{DAmbrosio:2021}.}. The $f(\Q)$ connection, on the other hand, becomes truly dynamical for both, black hole as well as cosmological spacetimes.
\\
The next natural step in the analysis of teleparallel cosmologies is the study of perturbations. That is, one chooses the ansatz  $g_{\mu\nu}=\mathring{g}_{\mu\nu}+\delta g_{\mu\nu}$ and $\Gamma^\alpha{}_{\mu\nu}=\mathring{\Gamma}^\alpha{}_{\mu\nu}+\delta\Gamma^\alpha{}_{\mu\nu}$, where the ring over $\mathring{g}$ and $\mathring{\Gamma}$ indicates the background solutions which solve the field equations presented in this work. One then expands these equations to first order in the perturbations $\delta g$ and $\delta \Gamma$. This leads to  equations for the perturbations, which can in principle be solved. However, the connection perturbations are subjected to the requirement of still satisfying the respective postulates of teleparallelism.  Together with the algebraically complicated field equations, the task of cosmological perturbation theory becomes a difficult one. For an attempt in the context of MT see for instance~\cite{Hohmann:2020}.

\section*{Acknowledgements}
LH is supported by funding from the European Research Council (ERC) under the European Unions Horizon 2020 research and innovation programme grant agreement No 801781 and by the Swiss National Science Foundation grant 179740.

\bibliographystyle{utcaps}
\bibliography{Bibliography}

\providecommand{\href}[2]{#2}\begingroup\raggedright\begin{thebibliography}{10}

\bibitem{BeltranJimenez:2019}
J.~B. Jim\'{e}nez, L.~Heisenberg, and T.~S. Koivisto, ``{The Geometrical
  Trinity of Gravity},'' \href{http://dx.doi.org/10.3390/universe5070173}{{\em
  Universe} {\bfseries 5} no.~7, (2019) 173},
  \href{http://arxiv.org/abs/1903.06830}{{\ttfamily arXiv:1903.06830
  [hep-th]}}.

\bibitem{Heisenberg:2018}
L.~Heisenberg, ``{A systematic approach to generalisations of General
  Relativity and their cosmological implications},''
  \href{http://dx.doi.org/10.1016/j.physrep.2018.11.006}{{\em Phys. Rept.}
  {\bfseries 796} (2019) 1--113},
  \href{http://arxiv.org/abs/1807.01725}{{\ttfamily arXiv:1807.01725 [gr-qc]}}.

\bibitem{Jimenez:2019}
J.~Beltr\'{a}n~Jim\'{e}nez, L.~Heisenberg, T.~S. Koivisto, and S.~Pekar,
  ``{Cosmology in $f(Q)$ geometry},''
  \href{http://dx.doi.org/10.1103/PhysRevD.101.103507}{{\em Phys. Rev. D}
  {\bfseries 101} no.~10, (2020) 103507},
  \href{http://arxiv.org/abs/1906.10027}{{\ttfamily arXiv:1906.10027 [gr-qc]}}.

\bibitem{Barros:2020}
B.~J. Barros, T.~Barreiro, T.~Koivisto, and N.~J. Nunes, ``{Testing $F(Q)$
  gravity with redshift space distortions},''
  \href{http://dx.doi.org/10.1016/j.dark.2020.100616}{{\em Phys. Dark Univ.}
  {\bfseries 30} (2020) 100616},
  \href{http://arxiv.org/abs/2004.07867}{{\ttfamily arXiv:2004.07867 [gr-qc]}}.

\bibitem{Ayuso:2020}
I.~Ayuso, R.~Lazkoz, and V.~Salzano, ``{Observational constraints on
  cosmological solutions of $f(Q)$ theories},''
  \href{http://dx.doi.org/10.1103/PhysRevD.103.063505}{{\em Phys. Rev. D}
  {\bfseries 103} no.~6, (2021) 063505},
  \href{http://arxiv.org/abs/2012.00046}{{\ttfamily arXiv:2012.00046
  [astro-ph.CO]}}.

\bibitem{Frusciante:2021}
N.~Frusciante, ``{Signatures of $f(Q)$-gravity in cosmology},''
  \href{http://dx.doi.org/10.1103/PhysRevD.103.044021}{{\em Phys. Rev. D}
  {\bfseries 103} no.~4, (2021) 044021},
  \href{http://arxiv.org/abs/2101.09242}{{\ttfamily arXiv:2101.09242
  [astro-ph.CO]}}.

\bibitem{Anagnostopoulos:2021}
F.~K. Anagnostopoulos, S.~Basilakos, and E.~N. Saridakis, ``{First evidence
  that non-metricity f(Q) gravity could challenge $\Lambda$CDM},''
  \href{http://arxiv.org/abs/2104.15123}{{\ttfamily arXiv:2104.15123 [gr-qc]}}.

\bibitem{Arora:2021}
S.~Arora, S.~K.~J. Pacif, A.~Parida, and P.~K. Sahoo, ``{Bulk viscous matter
  and the cosmic acceleration of the universe in $f(Q,T)$ gravity},''
  \href{http://arxiv.org/abs/2106.00491}{{\ttfamily arXiv:2106.00491 [gr-qc]}}.

\bibitem{Bahamonde:2021}
S.~Bahamonde, K.~F. Dialektopoulos, C.~Escamilla-Rivera, G.~Farrugia, V.~Gakis,
  M.~Hendry, M.~Hohmann, J.~L. Said, J.~Mifsud, and E.~Di~Valentino,
  ``{Teleparallel Gravity: From Theory to Cosmology},''
  \href{http://arxiv.org/abs/2106.13793}{{\ttfamily arXiv:2106.13793 [gr-qc]}}.

\bibitem{Esposito:2021}
F.~Esposito, S.~Carloni, R.~Cianci, and S.~Vignolo, ``{Recontructing isotropic
  and anisotropic $f(\mathcal{Q})$ cosmologies},''
  \href{http://arxiv.org/abs/2107.14522}{{\ttfamily arXiv:2107.14522 [gr-qc]}}.

\bibitem{Dimakis:2021}
N.~Dimakis, A.~Paliathanasis, and T.~Christodoulakis, ``{Quantum Cosmology in
  $f(Q)$ theory},'' \href{http://arxiv.org/abs/2108.01970}{{\ttfamily
  arXiv:2108.01970 [gr-qc]}}.

\bibitem{Atayde:2021}
L.~Atayde and N.~Frusciante, ``{Can $f(Q)$-gravity challenge $\Lambda$CDM?},''
  \href{http://arxiv.org/abs/2108.10832}{{\ttfamily arXiv:2108.10832
  [astro-ph.CO]}}.

\bibitem{Hohmann:2021}
M.~Hohmann, ``{General covariant symmetric teleparallel cosmology},''
  \href{http://arxiv.org/abs/2109.01525}{{\ttfamily arXiv:2109.01525 [gr-qc]}}.

\bibitem{DAmbrosio:2021}
F.~D'Ambrosio, S.~D.~B. Fell, L.~Heisenberg, and S.~Kuhn, ``{Black holes in
  $f(\mathbb Q)$ Gravity},'' \href{http://arxiv.org/abs/2109.03174}{{\ttfamily
  arXiv:2109.03174 [gr-qc]}}.

\bibitem{BeltranJimenez:2017}
J.~Beltr\'{a}n~Jim\'{e}nez, L.~Heisenberg, and T.~Koivisto, ``{Coincident
  General Relativity},''
  \href{http://dx.doi.org/10.1103/PhysRevD.98.044048}{{\em Phys. Rev. D}
  {\bfseries 98} no.~4, (2018) 044048},
  \href{http://arxiv.org/abs/1710.03116}{{\ttfamily arXiv:1710.03116 [gr-qc]}}.

\bibitem{Jimenez:2018}
J.~Beltr\'{a}n~Jim\'{e}nez, L.~Heisenberg, and T.~S. Koivisto, ``{Teleparallel
  Palatini theories},''
  \href{http://dx.doi.org/10.1088/1475-7516/2018/08/039}{{\em JCAP} {\bfseries
  1808} no.~08, (2018) 039},
\href{http://arxiv.org/abs/1803.10185}{{\ttfamily arXiv:1803.10185 [gr-qc]}}.

\bibitem{Hohmann:2019}
M.~Hohmann, ``{Metric-affine Geometries With Spherical Symmetry},''
  \href{http://dx.doi.org/10.3390/sym12030453}{{\em Symmetry} {\bfseries 12}
  no.~3, (2020) 453}, \href{http://arxiv.org/abs/1912.12906}{{\ttfamily
  arXiv:1912.12906 [math-ph]}}.

\bibitem{Hohmann:2019nat}
M.~Hohmann, L.~J\"arv, M.~Kr\v{s}\v{s}\'ak, and C.~Pfeifer, ``{Modified
  teleparallel theories of gravity in symmetric spacetimes},''
  \href{http://dx.doi.org/10.1103/PhysRevD.100.084002}{{\em Phys. Rev. D}
  {\bfseries 100} no.~8, (2019) 084002},
  \href{http://arxiv.org/abs/1901.05472}{{\ttfamily arXiv:1901.05472 [gr-qc]}}.

\bibitem{Mandal:2020}
S.~Mandal, D.~Wang, and P.~K. Sahoo, ``{Cosmography in $f(Q)$ gravity},''
  \href{http://dx.doi.org/10.1103/PhysRevD.102.124029}{{\em Phys. Rev. D}
  {\bfseries 102} (2020) 124029},
  \href{http://arxiv.org/abs/2011.00420}{{\ttfamily arXiv:2011.00420 [gr-qc]}}.

\bibitem{Hohmann:2020}
M.~Hohmann, ``{General cosmological perturbations in teleparallel gravity},''
  \href{http://dx.doi.org/10.1140/epjp/s13360-020-00969-6}{{\em Eur. Phys. J.
  Plus} {\bfseries 136} no.~1, (2021) 65},
  \href{http://arxiv.org/abs/2011.02491}{{\ttfamily arXiv:2011.02491 [gr-qc]}}.

\end{thebibliography}\endgroup

\end{document}